\pdfoutput=1
\documentclass[11pt,twoside,a4paper,cmspaper,final,collab]{cms-tdr}
\def\svnVersion{787c812}\def\svnDate{2024/09/09}\def\cmsCernNoTag{CERN-EP-2024-085}\def\cmsCernDate{\today}\def\cmsMessage{Published in Physics Letters B as \href{http://dx.doi.org/10.1016/j.physletb.2024.138979}{\doi{10.1016/j.physletb.2024.138979}.}}
\begin{document}\cmsNoteHeader{TOP-22-007}

\newlength\cmsTabSkip
\setlength\cmsTabSkip{1ex}

\ifthenelse{\boolean{cms@external}}{%
    \providecommand{\cmsLeft}{upper\xspace}
    \providecommand{\cmsRight}{lower\xspace}
    \providecommand{\cmsTable}[1]{\resizebox{0.9\textwidth}{!}{#1}}
    \providecommand{\cmsTableMaybe}[1]{\resizebox{\linewidth}{!}{#1}}
    \providecommand{\cmsTableNorm}[1]{#1}
}{%
    \providecommand{\cmsLeft}{left\xspace}
    \providecommand{\cmsRight}{right\xspace}
    \providecommand{\cmsTable}[1]{\resizebox{\textwidth}{!}{#1}}
    \providecommand{\cmsTableMaybe}[1]{#1}
    \providecommand{\cmsTableNorm}[1]{\cmsTable{#1}}
}

\newcommand{\mtop}{\ensuremath{m_{\PQt}}\xspace}
\newcommand{\pp}{\ensuremath{\Pp\Pp}\xspace}
\newcommand{\cmunu}{\ensuremath{c^{\mu\nu}}\xspace}
\newcommand{\dmunu}{\ensuremath{d^{\mu\nu}}\xspace}
\newcommand{\tW}{\ensuremath{\PQt\PW}\xspace}
\newcommand{\ttW}{\ensuremath{\ttbar\PW}\xspace}
\newcommand{\ttZ}{\ensuremath{\ttbar\PZ}\xspace}
\newcommand{\ttV}{\ensuremath{\ttbar\PV}\xspace}
\newcommand{\WW}{\ensuremath{\PW\PW}\xspace}
\newcommand{\WZ}{\ensuremath{\PW\PZ}\xspace}
\newcommand{\ZZ}{\ensuremath{\PZ\PZ}\xspace}
\newcommand{\Wjets}{\ensuremath{\PW\text{+jets}}\xspace}
\newcommand{\Zjets}{\ensuremath{\PZ\text{+jets}}\xspace}
\newcommand{\Vjets}{\ensuremath{\PV\text{+jets}}\xspace}
\newcommand{\hdamp}{\ensuremath{h_{\mathrm{damp}}}\xspace}
\newcommand{\ellell}{\ensuremath{\HepParticle{\Pell}{}{+}\HepParticle{\Pell}{}{-}}\xspace}
\newcommand{\sigavg}{\ensuremath{\sigma_{\mathrm{avg}}}\xspace}
\newcommand{\sigSM}{\ensuremath{\sigma_{\mathrm{SM}}}\xspace}
\newcommand{\sigSME}{\ensuremath{\sigma_{\mathrm{SME}}}\xspace}
\newcommand{\muavg}{\ensuremath{\mu_{\mathrm{avg}}}\xspace}
\newcommand{\pers}{\ensuremath{\unit{s}^{-1}}\xspace}
\newcommand{\Omgsid}{\ensuremath{\Omega_{\text{sidereal}}}\xspace}
\newcommand{\OmgUTC}{\ensuremath{\Omega_{\text{UTC}}}\xspace}
\newcommand{\tUNIX}{\ensuremath{t_{\text{UNIX}}}\xspace}
\newcommand{\phiUNIX}{\ensuremath{\phi_{\text{UNIX}}}\xspace}
\newcommand{\philong}{\ensuremath{\phi_{\text{longitude}}}\xspace}
\newcommand{\cLmunu}{\ensuremath{c_{L,\mu\nu}}\xspace}
\newcommand{\cRmunu}{\ensuremath{c_{R,\mu\nu}}\xspace}

\cmsNoteHeader{TOP-22-007}
\title{Searches for violation of Lorentz invariance in top quark pair production using dilepton events in \texorpdfstring{13\TeV}{13 TeV} proton-proton collisions}

\author{CMS Collaboration}
\date{\today}

\abstract{A search for violation of Lorentz invariance in the production of top quark pairs (\ttbar) is presented. The measured normalized differential \ttbar production cross section, as a function of the sidereal time, is examined for potential modulations induced by Lorentz-invariance breaking operators in an effective field theory extension of the standard model (SM). The cross section is measured from collision events collected by the CMS detector at a center-of-mass-energy of 13\TeV, corresponding to an integrated luminosity of 77.8\fbinv, and containing one electron and one muon. The results are found to be compatible with zero, in agreement with the SM, and are used to place upper limits at 68\% confidence level on the magnitude of the Lorentz-violating couplings ranging from 1--$8\times 10^{-3}$. This is the first precision test of the isotropy in special relativity with top quarks at the LHC, restricting further the bounds on such couplings by up to two orders of magnitude with respect to previous searches conducted at the Tevatron.}

\hypersetup{%
pdfauthor={CMS Collaboration},%
pdftitle={Searches for violation of Lorentz invariance in top quark pair production using dilepton events in 13 TeV proton-proton collisions},%
pdfsubject={CMS},%
pdfkeywords={CMS, top quark, Beyond Standard Model, Lorentz symmetry}}

\maketitle

\section{Introduction}

Lorentz invariance is a key ingredient of the relativistic description of our world, on the scale of elementary particles as well as on large scales in the Universe. The standard model (SM) of particle physics is a relativistic quantum field theory, and as such it must contain a Lorentz-invariant action. General relativity preserves Lorentz invariance locally.
The Lorentz group contains rotations and boosts in Minkowski spacetime.
Models of quantum gravity, \eg, string theory~\cite{Kostelecky:1988zi} and loop quantum gravity~\cite{Gambini:1998it}, can, however, predict breaking of Lorentz invariance at a high mass scale.
A quantum theory including gravity may be governed by an energy scale lower than the Planck mass, possibly at the \TeVns scale~\cite{Pospelov:2013mpa}.
Thus, induced deviations from Lorentz invariance may be measurable at the LHC.

The standard model extension (SME)~\cite{Colladay:1996iz, Colladay:1998fq} is an effective field theory in which all operators violating Lorentz invariance are added to the SM Lagrangian.
Its coefficients, controlling the size of Lorentz invariance breaking, are different for each particle.
While coefficients related to photons, neutrons, protons, and neutrinos have been measured precisely~\cite{Kostelecky:2008ts}, the quark sector remains constrained at a relatively lower precision.
The ZEUS Collaboration at the DESY HERA recently searched for violation of Lorentz invariance arising from light quarks in deep inelastic scattering data~\cite{ZEUS:2022msi}.
Searches for Lorentz invariance violation were performed in neutral meson mixing at KLOE~\cite{KLOE-2:2013ozx}, KTeV~\cite{Nguyen:2001tg}, FOCUS~\cite{Link:2002fg}, BaBar~\cite{BaBar:2007run}, \DZERO~\cite{D0:2015ycz}, and LHCb~\cite{Aaij:2016mos}.
Violation of Lorentz invariance was searched for with top quark pairs (\ttbar) by the \DZERO experiment~\cite{Abazov:2012iu} at the Fermilab Tevatron, showing compatibility with Lorentz invariance with an absolute uncertainty of about 10\% on the tested SME coefficients.
The proton-proton (\pp) collisions at the CERN LHC at $\sqrt{s}=13\TeV$ produce \ttbar events with a cross section approximately 100 times higher than that of the Tevatron.
Studies showed that the measurements at the LHC have a large potential to significantly improve upon existing results~\cite{Carle:2019ouy}.

In this Letter, the first search for Lorentz invariance violation with top quarks at the LHC is presented, within the context of the SME.
Signatures for violation of Lorentz invariance involving top quarks are parameterized with the SME Lagrangian~\cite{Berger:2015yha}:
\begin{equation}
\label{eqFt1}
\mathcal{L}_{\mathrm{SME}} = \frac{1}{2} i \bar{\psi} ( \gamma^{\nu} + c^{\mu\nu} \gamma_{\mu}  + d^{\mu\nu} \gamma_5 \gamma_{\mu} ) \overleftrightarrow{\partial_{\nu}} \psi - \mtop \bar{\psi} \psi,
\end{equation}
where $\psi$ and $\bar{\psi}$ are the Dirac fields for top quarks and antiquarks, \mtop is the top quark mass, and \cmunu and \dmunu are the unitless SME coefficients. Unlike the SM fields, \cmunu and \dmunu are not modified under Lorentz transformations of the particle system:\ they are constant $4{\times}4$ matrices, thereby breaking Lorentz invariance of the Lagrangian. They indicate a preferential direction in spacetime as seen by top quarks, violating the isotropy of special relativity.
Boosts associated with the Earth's rotation and with the revolution around the sun are negligible relative to top quark boosts.
Lorentz invariance is preserved when \cmunu and \dmunu are zero.

Choosing a reference frame is needed to report measurements of Lorentz-violating coefficients,
in analogy, for instance, with the measurement of particle energies, which are not invariant under Lorentz transformation.
By convention, results are reported in the sun-centered frame (SCF)~\cite{Kostelecky:2008ts}, with its origin at the center of the sun; the $Z$ axis pointing north, parallel to the Earth's rotation axis; the $X$ axis pointing to the intersection of the ecliptic and celestial equator on January 1st, 2000 (J2000); and the $Y$ axis completing the direct basis.
The SCF can be considered as inertial in the lifetime of a physics experiment.
Hereafter, the coefficients \cmunu and \dmunu with space-time indices $\mu$ or $\nu=T,X,Y,Z$ are expressed in the SCF (to be distinguished from the space-time indices of the CMS frame).
The relevant measure of time in this reference frame is called sidereal time.
While one rotation period of the Earth is equal to approximately 23\unit{h} 56\unit{min} UTC (\ie, Universal Time Coordinated, where UTC seconds are defined from Cesium atomic hyperfine transitions), one rotation period is defined as being equal to 24 sidereal hours.
The CMS detector moves around the Earth's rotation axis during a sidereal day, and so does the beam line direction at the interaction point, or the average direction of top quarks produced in the collisions.
As a consequence, top quark couplings \cmunu and \dmunu depend on time, resulting in cross sections for top quark production modulating with sidereal time. Searching for such signatures is the goal of this Letter.

We report a search for violation of Lorentz invariance in $\pp \to \ttbar$ production with the CMS experiment at $\sqrt{s}=13\TeV$, using events with one muon and one electron of opposite charge in the final state, stemming from the decay $\ttbar\to\bbbar\Pe\PGne\PGm\PGnGm$ (where the particle or antiparticle nature of the leptons is implied).
The $\Pe\PGm$ channel provides a high purity sample of \ttbar events owing to the negligible contribution from \Zjets processes.
A data sample recorded in 2016--2017 with the CMS experiment and corresponding to an integrated luminosity of 77.8\fbinv is analyzed.
Events are categorized based on the number of jets identified as originating from \PQb quarks (\PQb jets) to improve the separation between the \ttbar signal and the background processes. The normalized \ttbar cross section is measured differentially as a function of the sidereal time, in bins of hours within the sidereal day, by performing a profile maximum likelihood fit to the event yields in the \PQb jet multiplicity. The measurement is used to constrain, for the first time at the LHC, the Lorentz-violating coefficients \cmunu and \dmunu impacting top quarks.
Tabulated results are provided in the HEPData record for this analysis~\cite{hepdata}.

The Letter is organized as follows. A description of the CMS detector and event reconstruction is provided in Section~\ref{reco}. Section~\ref{Samples} gives details on simulation samples, and Section~\ref{selection} describes the event selection. Event corrections and associated systematic uncertainties, as well as their treatment relative to sidereal time, are detailed in Section~\ref{Systematics}. The normalized differential cross section for \ttbar production as a function of sidereal time is measured in Section~\ref{DifferentialMeas}, providing a model-independent result. Bounds on the coefficients \cmunu and \dmunu within the SME framework are extracted in Section~\ref{SignalExtraction}. The Letter concludes with a Summary in Section~\ref{Conclusions}.

\section{\label{reco}Event reconstruction with the CMS detector}

The CMS apparatus~\cite{CMS:Detector-2008, CMSTrackerGroup:2020edz, CMS:PRF-21-001} is a multipurpose, nearly hermetic detector, designed to trigger on~\cite{CMS:TRG-17-001,CMS:TRG-12-001} and identify electrons, muons, photons, and hadrons~\cite{CMS:EGM-17-001,CMS:MUO-16-001,CMS:TRK-11-001}.
A global ``particle-flow'' (PF) algorithm~\cite{CMS:PRF-14-001} aims to reconstruct all individual particles in an event, combining information provided by the all-silicon inner tracker and by the crystal electromagnetic (ECAL) and brass-scintillator hadron (HCAL) calorimeters, operating inside a 3.8\unit{T} superconducting solenoid, with data from the gas-ionization muon detectors embedded in the flux-return yoke outside the solenoid.
The reconstructed particles are used to build \PGt leptons, jets, and missing transverse momentum~\cite{CMS:TAU-16-003,CMS:JME-13-004,CMS:JME-17-001}.
A more detailed description of the CMS detector, together with a definition of the coordinate system used and the relevant kinematic variables, can be found in Refs.~\cite{CMS:Detector-2008, CMSTrackerGroup:2020edz, CMS:PRF-21-001}.

Events of interest are selected using a two-tiered trigger system. The first level (L1), composed of custom hardware processors, uses information from the calorimeters and muon detectors to select events at a rate of around 100\unit{kHz} within a fixed latency of about 4\mus~\cite{CMS:TRG-17-001}.
During the 2016--2017 data taking, a gradual shift in the timing of the inputs of the ECAL L1 trigger in the region at $\abs{\eta} > 2.0$, where $\eta$ is the pseudorapidity, caused a specific trigger inefficiency. Correction factors, which are small, were computed from data and applied to the simulation.
The second level, known as the high-level trigger (HLT), consists of a farm of processors running a version of the full event reconstruction software optimized for fast processing, and reduces the event rate to around 1\unit{kHz} before data storage~\cite{CMS:TRG-12-001}.
Events are selected at the HLT by requiring the presence of one or two leptons, with transverse momentum (\pt) thresholds as follows:\ one electron with $\pt > 12\GeV$ and one muon with $\pt>23\GeV$; one electron with $\pt > 23\GeV$ and one muon with $\pt>8\GeV$; one electron with $\pt > 27$ (35)\GeV in 2016 (2017); or one muon with $\pt > 24$ (27)\GeV in 2016 (2017).

The electron momentum is estimated by combining the energy measurement in the ECAL with the momentum measurement in the tracker.
Electron candidates are required to satisfy $\pt > 25$ and 20\GeV for the highest \pt (leading) and second-highest \pt (subleading) lepton. Furthermore, electrons must have $\abs{\eta} < 2.4$, and their energy deposit in the ECAL must lie outside of the transition region between the barrel and endcap sections ($1.44 < \abs{\eta} <1.56$).
Electron candidates must satisfy tight identification criteria and are required to be isolated in the detector to reject jets misidentified as electrons.
Identification criteria comprise selections on a shower shape variable in the ECAL, on the matching of the track with the ECAL energy deposit, on the ratio of HCAL over ECAL energy deposits around the candidate, the hits in the inner tracker, and on criteria to reject electrons arising from photon conversions~\cite{CMS:EGM-17-001}.
An isolation variable is computed with the PF algorithm as the scalar \pt sum of photons, charged and neutral hadrons in a cone around the electron direction. The cone is defined by $\DR=\sqrt{\smash[b]{(\Delta\eta)^2+(\Delta\phi)^2}} < 0.3$, where $\Delta\eta$ and $\Delta\phi$ are the $\eta$ and azimuthal angle differences between the PF particles and the electron, respectively. Contributions to the isolation sum from other inelastic \pp interactions in the same or nearby bunch crossing (pileup) are corrected using the \FASTJET estimator~\cite{EXT:FastJet-2012}.
The relative isolation, calculated as the isolation sum divided by the electron \pt, is used.
Differences in reconstruction and identification efficiencies between data and simulation are corrected using $\PZ\to\EE$ events with a ``tag-and-probe'' method~\cite{CMS:EWK-10-002}.

Muons are measured in the range $\abs{\eta} < 2.4$, with detection planes made using three technologies:\ drift tubes, cathode strip chambers, and resistive-plate chambers.
Muon candidates are reconstructed with the PF algorithm. They are required to satisfy $\pt > 25$ (20)\GeV for the leading (subleading) lepton.
Muon candidates are required to satisfy tight identification and isolation criteria, to suppress contributions from jets misidentified as muons.
Identification criteria include selections on the number of matched muon detector planes, hits in the pixel and tracker, the $\chi^2$ of the trajectory fit including tracking layers and muon detector planes, and the transverse impact parameter of the track~\cite{CMS:MUO-16-001}.
A relative isolation variable is calculated with the PF algorithm similar to electrons but with a cone size $\DR<0.4$.
Differences in identification and isolation efficiencies between data and simulation are corrected using $\PZ\to\MM$ events with the tag-and-probe method.

Jets are reconstructed from PF candidates, using the anti-\kt algorithm with a distance parameter of 0.4~\cite{EXT:AK-2008, EXT:FastJet-2012}.
Jet momentum is determined as the vectorial sum of all particle momenta in the jet, and is found from simulation to be, on average, within 5--10\% of the true momentum over the entire \pt spectrum and detector acceptance.
To mitigate pileup contributions, tracks identified to be originating from pileup vertices are discarded, and an offset correction is applied to correct for the remaining contributions.
Jet energy corrections are derived from simulation studies so that the average measured energy of jets becomes identical to that of particle level jets.
In situ measurements of the momentum balance in dijet, photon+jet, \PZ{}+jet, and multijet events are used to determine any residual differences between the jet energy scale in data and simulation, and appropriate corrections are made~\cite{CMS:JME-13-004}.
Additional selection criteria are applied to each jet to remove jets potentially dominated by instrumental effects or reconstruction failures.
Jets are selected if they satisfy $\pt > 30\GeV$ and $\abs{\eta}<2.4$.
Jets are not considered if their distance \DR from selected leptons is less than 0.4.

For the identification of \PQb jets, a loose selection on the score of the \textsc{deepCSV} algorithm, a deep-learning discriminant using secondary vertex and lifetime information~\cite{CMS:BTV-16-002, CMS:DP-2023-005}, is used.
The identification efficiency of \PQb quark jets is about 84\% with a probability of misidentifying light-quark and gluon jets of 10\%, estimated for \PQb jet candidates with $\pt > 20\GeV$ in simulated \ttbar events.
Corrections for the difference in identification efficiency of \PQb jets and light jets between data and simulation are computed as a function of the jet \pt and $\eta$ and applied to all jets in the event.

\section{\label{Samples}Simulation samples}

Monte Carlo (MC) samples for \ttbar production are generated at next-to-leading order (NLO) in perturbative quantum chromodynamics (pQCD) with the \POWHEG v2~\cite{EXT:Powheg-2010} generator, interfaced with \PYTHIA8~\cite{Sjostrand:2007gs} for simulating parton shower, hadronization, and the underlying event.
Revision 3453 is used for all the MC samples generated with \POWHEG, while all of the simulation samples use \PYTHIA8.226 (8.230) in 2016 (2017).
The \ttbar cross section is computed with the \textsc{top++} v2.0 program~\cite{Czakon:2011xx} at next-to-NLO (NNLO) and next-to-next-to-leading logarithmic accuracy.
It has been observed that the top quark \pt spectrum in data is softer than the spectrum predicted in simulation~\cite{CMS:TOP-12-028,CMS:TOP-16-008,CMS:TOP-20-006}. Simulated events for \ttbar production are therefore corrected for the ratio of unfolded \ttbar data over \POWHEG predictions as a function of the top quark and antiquark \pt at parton level.

Since this analysis targets the \ttbar final state with two charged leptons of different flavors and at least one \PQb jet, single top quark production in association with a \PW boson (\tW) constitutes the main background, with one lepton each arising from the top quark and associated \PW boson decay.
Simulated samples of \tW and $t$-channel single top quark production are generated at NLO in pQCD with \POWHEG v2.
The cross section for the \tW channel is computed at approximate NNLO~\cite{Kidonakis:2021vob}.
Other backgrounds involving top quarks include $s$-channel single top quark production and \ttbar production in association with a vector boson (\ttW and \ttZ, grouped together and denoted as \ttV). Simulated samples for these processes are generated at NLO with \MGvATNLO~\cite{EXT:MG5aMCatNLO-2014} (version 2.4.2 in 2016 and 2.6.0 in 2017 in all samples using \MGvATNLO) with additional jets using FxFx merging~\cite{EXT:FxFx-2012}.
The cross sections for single top $t$- and $s$-channel production are computed with the \textsc{hathor} v2.1 program~\cite{Kant:2014oha} at NLO accuracy.
The cross sections for \ttW and \ttZ processes are computed at NLO with \MGvATNLO generation~\cite{EXT:MG5aMCatNLO-2014}.
In the simulated samples, a fixed value of $\mtop = 172.5\GeV$ is used.
Diboson production samples (\WW, \WZ, \ZZ) are generated with \PYTHIA8.
The cross section for the \WW process is computed at NNLO~\cite{Gehrmann:2014fva}, while the \WZ and \ZZ cross sections are computed at NLO with \MCFM~\cite{Campbell:2011bn} version 6.6.
The \Wjets and \Zjets processes (grouped together and denoted as \Vjets) are generated using \MGvATNLO, at leading order (LO) matched with additional jets using the MLM algorithm~\cite{Alwall:2008qv} or at NLO with additional jets using FxFx merging.
Cross sections for \Zjets and \Wjets processes are computed with \FEWZ~\cite{Gavin:2010az} version 3.1 at NNLO.

The underlying event is modeled with the CP5 tune~\cite{CMS:GEN-17-001} for all of the simulated samples with 2017 detector conditions, and for the simulation of \ttbar and \tW processes with 2016 detector conditions.
The samples simulated with 2016 detector conditions are using the CUETP8M1 tune~\cite{CMS:GEN-14-001} for all other background processes.
The parton distribution functions (PDFs) used are the NNPDF3.1 NNLO set~\cite{EXT:NNPDF-2017} for samples using the CP5 tune, and NNPDF3.0 NNLO for samples using the CUETP8M1 tune.

The CMS detector response is simulated with the \GEANTfour toolkit~\cite{EXT:GEANT4-2002}.
Data and simulation events are passed through the same CMS software reconstruction chain.
Pileup interactions generated with \PYTHIA8 are added to the simulations, and the distribution in the number of pileup interactions is corrected for each MC sample to reproduce the numbers observed in data.

\section{\label{selection}Event selection}

Events with exactly one electron and one muon are selected.
Events are rejected if there are additional electrons or muons with $\pt > 10\GeV$ and identified with looser criteria.
In addition, the invariant mass of the dilepton pair is required to satisfy $m_{\Pe\PGm} > 20\GeV$, to reduce contributions from low-mass resonances and Drell--Yan \PGt pair production.
Events with at least two jets are selected, among which at least one must be identified as a \PQb jet.

\begin{figure}[ht!]
\centering
\includegraphics[width=0.47\textwidth]{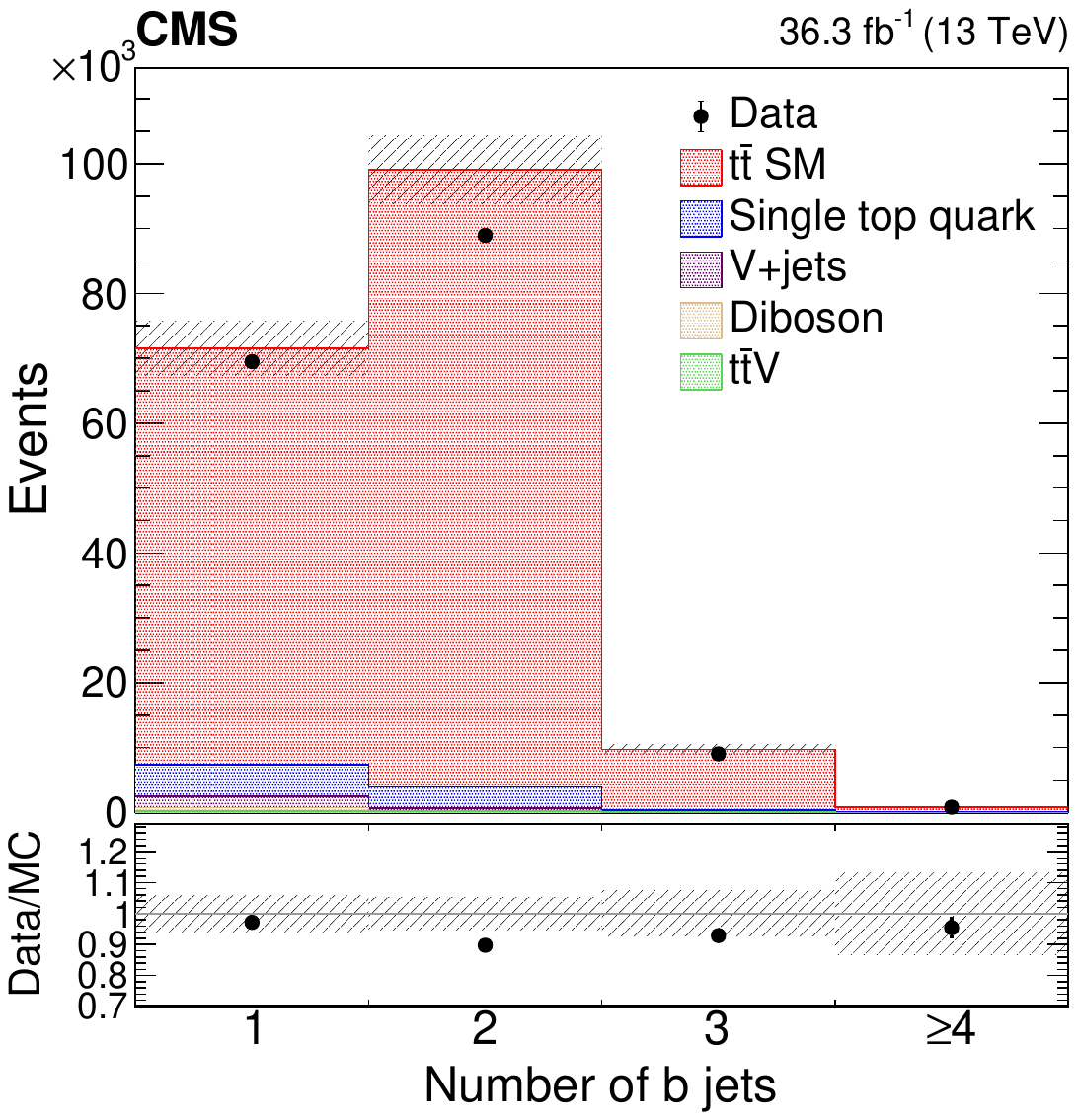}
\includegraphics[width=0.47\textwidth]{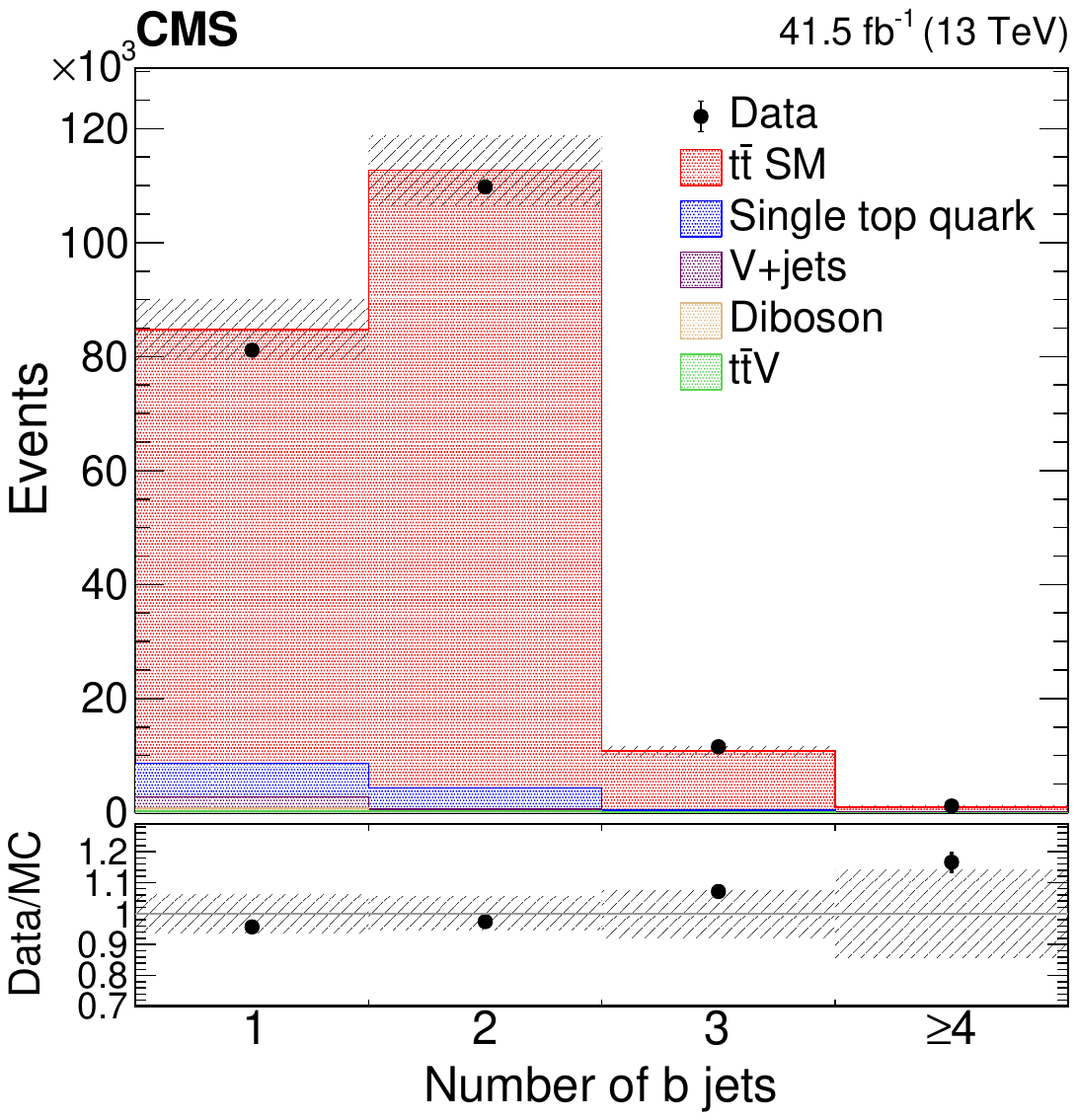}
\caption{\label{fig:DataMCcomp_nbjets}Distribution of the number of \PQb jets in data and simulation, after the event selection, (\cmsLeft) in 2016 and (\cmsRight) in 2017 samples. The hatched band includes statistical and systematic uncertainties in the predictions. The vertical bars associated with the data points represent their statistical uncertainty. The lower panels show the ratio of the observed data event yields to those expected from simulation.}
\end{figure}

The distribution in the number of \PQb jets is compared between data and simulation in Fig.~\ref{fig:DataMCcomp_nbjets}.
The MC simulation of signal and backgrounds provides a description of the data distribution in the number of \PQb jets agreeing within uncertainties.
Inclusively, the event yields are 7.1\% (2016) and 2.7\% (2017) lower in data than in simulation, as shown in Table~\ref{table:DataMCcomp_inclusive}.

\begin{table}[ht!]
\centering\renewcommand\arraystretch{1.1}
\topcaption{\label{table:DataMCcomp_inclusive}Event yields in data and MC simulation in 2016--2017, after selection. The uncertainties include statistical and systematic sources, with correlations.}
\cmsTableMaybe{\begin{tabular}{cr@{${}\pm{}$}lr@{${}\pm{}$}l}
Process & \multicolumn{2}{c}{2016} & \multicolumn{2}{c}{2017} \\
\hline
\ttbar SM & 169\,500 & 6100 & 196\,000 & 7300 \\
\ttV & 460 & 60 & 540 & 70 \\
Single top quark & 8\,500 & 1600 & 9\,900 & 2000 \\
Diboson & 700 & 150 & 650 & 130 \\
\Vjets & 2\,100 & 500 & 2\,300 & 600 \\[\cmsTabSkip]
Total background & 11\,700 & 1700 & 13\,404 & 2000 \\
Total MC & 181\,000 & 6700 & 209\,000 & 8100 \\[\cmsTabSkip]
Data & \multicolumn{2}{c}{168\,282} & \multicolumn{2}{c}{203\,584} \\
\end{tabular}}
\label{tab:data1}
\end{table}

The distribution in the number of \PQb jets is used as discriminant for the signal extraction in Sections~\ref{DifferentialMeas} and~\ref{SignalExtraction}, since it provides clear separation between the \ttbar signal and the single top quark background, with the latter having only one top quark decaying to $\PQb\PW$, compared to two for the signal process.

\section{\label{Systematics}Treatment of the sidereal time dependence in corrections and systematic uncertainties}

The CMS data is recorded in UNIX time, whose seconds having the same duration as UTC seconds. The UNIX timestamp of the events is translated to sidereal time with the following formula:
\begin{equation}
\Omgsid\,t_{\mathrm{sidereal}} = \OmgUTC\,\tUNIX + \phiUNIX + \philong,
\end{equation}
where $\Omgsid = 2\pi / 86\,400\pers$ (sidereal) is the angular velocity of Earth's rotation around its axis in sidereal time, $\OmgUTC = 2\pi / 86\,164\pers$ (UTC) is the angular velocity of Earth's rotation around its axis in UTC time, and \tUNIX represents the event timestamp recorded at CMS in UNIX time since  January 1st, 1970 (UNIX epoch).
The azimuthal angle \phiUNIX encodes the phase between the UNIX epoch and J2000, which is the origin of the sidereal time count, and \philong is the effective longitude, defined as the angle between the the LHC beam at the CMS interaction point and the Greenwich meridian in rad.

Data events are classified in 24 bins, corresponding to the 24 sidereal hours within a sidereal day.
In the MC simulation, event timestamps are not included; instead, a time-integrated distribution is built and scaled by $1/24$ to define the expected prediction in each sidereal hour bin.
This prediction and its associated systematic uncertainties need to be corrected for effects that would potentially depend on sidereal time.
According to the definition of sidereal time, events recorded every day within a year at the same UTC time would populate the bins uniformly.

The instantaneous luminosity delivered by the LHC, and thus the distribution of the number of pileup interactions per bunch crossing evolves over the year.
To allow for differences in the instantaneous luminosity profile between sidereal time bins, the distribution in the number of pileup interactions in simulation is corrected to agree with data, as a function of the sidereal hour within a day.
The associated uncertainty arises from our imprecise knowledge of the total inelastic cross section, and is computed by varying this cross section by its uncertainty (4.6\%) simultaneously in each sidereal time bin.

The recorded integrated luminosity per sidereal time bin varies by up to 25\% within a sidereal day, reflecting the time structure of the LHC operations. The simulated events are corrected to match the integrated luminosity in each sidereal time bin.
The uncertainty in the integrated luminosity measurement originates from various sources in the normalization (related to the van der Meer calibration procedure~\cite{vanderMeer:296752, Grafstrom:2015foa, CMS:LUM-17-003}) and the integration of the luminosity measurement performed with several independent subdetectors (referred to as luminometers).
The normalization, as well as detector-specific normalization uncertainties are not expected to depend on sidereal time and are treated as uniform.
The uncertainties in the time stability and the linearity with respect to pileup, however, are estimated from cross-detector comparisons, \ie, by comparing the agreement between different luminometers, and their dependence on sidereal time can be evaluated explicitly.
In 2016, the three main luminometers and methods used for the luminosity measurement~\cite{CMS:LUM-17-003} are the cluster counting in the pixel tracker, an occupancy-based algorithm using the HCAL measurement, and an LHC radiation monitoring system.
In 2017, the three main luminometers and methods~\cite{CMS:LUM-17-004} are a method based on the sum of transverse energy in HCAL towers, pixel detector cluster counting, and the pixel luminosity telescope~\cite{CMS:NOTE-2022-007}.
The main luminometers used for the luminosity measurement are not the same at all times, since the individual luminosity detectors can be affected by various operational issues.
A small dependence of the stability and linearity estimates on the sidereal time is found:
\begin{itemize}
\item The cross-detector stability is evaluated separately in each sidereal time bin by taking the maximum difference among the integrated luminosities estimated with the three main luminometers. The relative uncertainty is found to be 0.1--0.3 (0.2--0.4)\% in 2016 (2017), consistent with the time-integrated uncertainties evaluated in Refs.~\cite{CMS:LUM-17-004,CMS:LUM-17-003}.
{\tolerance=800
\item The time dependence of the linearity of the luminometer response is evaluated in each sidereal time bin by computing the average single-bunch instantaneous luminosity (SBIL), as measured with the different luminometers, and weighting the largest difference with the corresponding cross-detector linearity slopes from Refs.~\cite{CMS:LUM-17-004,CMS:LUM-17-003}. The slopes are obtained from a fit of the ratios of SBIL estimates for pairs of luminometers, as a function of SBIL. The relative uncertainty is found to be at most 0.3 (0.5)\% in 2016 (2017), depending on the sidereal time bin.  The 2016 result is in agreement with the time-integrated uncertainty evaluated in Ref.~\cite{CMS:LUM-17-003}, but the 2017 result is smaller than the uncertainty from Ref.~\cite{CMS:LUM-17-004} due to the usage of the less-conservative approach from Ref.~\cite{CMS:LUM-17-003} for the evaluation of the 2017 luminosity.
\par}
\end{itemize}
Cross-detector stability and linearity uncertainties are included in the analysis with explicit dependence on sidereal time.
The LHC-related uncertainties (normalization uncertainties related to the van der Meer scan procedure) are evaluated to be 0.6 (0.9)\% in 2016 (2017), and other luminosity uncertainties related to the detector are evaluated to be 0.9 (1.4)\% in 2016 (2017).

The trigger efficiency, potentially changing slightly over the course of the year with the instantaneous luminosity profile, might also depend on sidereal time.
Simulated events are therefore corrected for the difference in the trigger efficiency between data and simulation as a function of sidereal time.
The trigger efficiencies are measured with the same method as in Ref.~\cite{CMS:TOP-19-006}, using events triggered with requirements on the missing transverse momentum \ptmiss, which is the magnitude of the \ptvecmiss vector,
defined as the projection onto the plane perpendicular to the beam axis of the negative vector sum of the momenta of all PF objects in the event, corrected for pileup effects.
Events triggered with \ptmiss requirements are correlated at less than 1\% with those passing single-lepton and dilepton triggers.
Events must satisfy a requirement on reconstructed $\ptmiss > 100\GeV$.
At least one jet identified as \PQb jet is required.
For each sidereal time bin, a statistical uncertainty is computed, and two systematic uncertainties are evaluated by repeating the measurement in subpartitions of the data:\ an uncertainty for the dependence of the trigger efficiency on the number of jets and on the period of data taking.
The ratio of trigger efficiencies in data and simulation is found to be in a range 97--99\%, with uncertainties at the level of 1\%, depending on the sidereal time bin.

For other experimental corrections and uncertainties, the dependence on sidereal time is unknown.
As a consequence, an individual nuisance parameter per sidereal time bin is conservatively attributed to most of the other experimental uncertainties, whose size is evaluated from an estimate not depending on time, allowing the fit to accommodate possible time dependence.
These experimental uncertainties include the following sources:\ ratios of efficiencies between data and simulation for electron and muon reconstruction, identification, and isolation that arise from tag-and-probe measurements with $\PZ\to\ellell$ events~\cite{CMS:EGM-17-001,CMS:MUO-16-001} (where $\Pell=\Pe$ or \PGm); uncertainties associated with the jet energy corrections (including jet energy resolution)~\cite{CMS:JME-13-004}; uncertainties arising from the identification of \PQb jets~\cite{CMS:BTV-16-002}; and uncertainties associated with the gradual shift in the timing of the inputs of the ECAL L1 trigger.
All of these uncertainties are treated as impacting the shape of the number of \PQb jets distribution, as well as its normalization, and are treated as uncorrelated across sidereal time.
Where relevant, each uncertainty source is split into systematic uncertainties correlated year-to-year, and a year-to-year uncorrelated component essentially of statistical nature.
Two additional sources of theoretical origin impact experimental quantities and are treated as correlated as functions of sidereal time and years:\ the uncertainty in the jet energy scale arising from the parton flavors, obtained by comparing \HERWIG and \PYTHIA simulation; and the uncertainty in lepton isolation arising from the phase space extrapolation of efficiencies, from $\PZ\to\ellell$ to \ttbar events.
The uncertainties inducing the largest impact on the distribution of the number of \PQb jets for the \ttbar process are the electron identification (1.2\% in 2016 and 2.1\% in 2017) and \PQb jet identification uncertainty (2\% at low number of jets to 5\% at high number of jets).

Contributions from background processes are estimated from simulation. After the selection, the event yield for background processes is about 8\% of the total number of expected events (Table~\ref{table:DataMCcomp_inclusive}), dominated by single top quark processes. The normalization of single top quark processes, whose contribution is dominated by the \tW process, is assigned an uncertainty of 30\%, accounting for the interference with the \ttbar process in the selected phase space. Diboson and \Vjets processes are attributed an uncertainty of 30\% as well estimated for the predictions in the selected phase-space. All these uncertainties are estimated following the prescription from Ref.~\cite{CMS:TOP-17-001}. An uncertainty of 20\% is associated with \ttV processes~\cite{CMS:TOP-17-005}. A normalization uncertainty of 4\% is assigned to the SM \ttbar process~\cite{CMS:TOP-17-001} in the SME fit (Section~\ref{SignalExtraction}).

The uncertainty corresponding to the correction of the top quark \pt in the \ttbar process is estimated by the difference with respect to not applying the correction (signatures for violation of Lorentz invariance searched for in this Letter cannot induce a modification of the top quark and antiquark \pt, once integrated over time).
Other theoretical uncertainties are arising from the top quark mass (obtained by varying \mtop by 1\GeV up and down around its nominal value); from the parameter \hdamp controlling the matching between matrix element and parton shower in the \POWHEG generator (varied around its nominal value as $1.379^{+0.926}_{-0.505}$ times \mtop); uncertainties associated with the CP5 underlying event tune (estimated during the tuning process~\cite{CMS:GEN-17-001}); and color reconnection uncertainties in the parton shower.
Uncertainties in the parton shower modeling, related to initial- and final-state radiation, are included for \ttbar and single top quark processes by varying the renormalization scale for initial- and final-state QCD emission in the parton shower by a factor 2 up and down.
Uncertainties arising from our missing knowledge of higher-order corrections in pQCD are included separately for \ttbar, single top quark, \ttV, and \Vjets processes by constructing an envelope of maximum variation computed from renormalization and factorization scale varied by a factor 1/2 to 2, excluding anticorrelated variations.
Uncertainties in the PDFs and the strong coupling constant \alpS are evaluated with the Hessian set of NNPDF3.1 (102 replicas) and added in quadrature, with a single nuisance parameter for \ttbar, single top quark, \ttV, and \Vjets processes.
The \hdamp, CP5 tune, and color reconnection sources modify the normalization of the distribution in the number of \PQb jets, while all the other theoretical uncertainties modify the shape of the distribution.
Theoretical uncertainties related to SM predictions are uniform across sidereal time.

A theoretical uncertainty corresponding to single top quark time modulation in the SME is included in the SME fit (described in Section~\ref{SignalExtraction}), and treated as correlated across sidereal time.

The same MC samples are used to predict the event yields in each sidereal time bin, separately for 2016 and 2017.
Statistical uncertainties in the MC samples are included as log-normal uncertainties independently for each bin of the \PQb jet multiplicity distribution, treated as correlated across sidereal time bins.

The assumed correlations of the uncertainties between years of data taking and sidereal time bins are detailed in Table~\ref{CorrelationScheme}.

\begin{table*}[p!]
\centering
\topcaption{\label{CorrelationScheme}Summary of the systematic uncertainties and their correlation scheme between 2016 and 2017 data sets, and between sidereal time bins. Sources marked with an asterisk are only included in the SME fits. Sources marked with a dagger are uniform and correlated in sidereal time.}
\renewcommand{\arraystretch}{1.1}
\cmsTableNorm{\begin{tabular}{cccc}
\multirow{2}{*}{Systematic uncertainty source} & Correlation & Correlation & \multirow{2}{*}{Magnitude} \\
& 2016--2017 & time bins & \\
\hline
Flat luminosity, year-to-year correlated part  & 100\% & 100\% & 0.6--0.9\%  \\
Flat luminosity, year-to-year uncorrelated part & 0\% & 100\% & 0.9--1.4\%  \\
Time-dependent luminosity stability & 0\% & 100\% & 0.2--0.4\% \\
Time-dependent luminosity linearity & 0\% & 100\% & 0.2--0.4\%  \\
Time-dependent pileup reweighting & 100\% & 100\% & 0.3--5\% \\
Time-dependent trigger efficiency, syst. component & 0\% & 100\% & 0.5--1\% \\
Time-dependent trigger efficiency, stat. component & 0\% & 0\% & 0.5\% \\[\cmsTabSkip]
L1 ECAL timing shift & 100\% & 0\% & 0.5\% \\
Electron reconstruction  & 100\% & 0\% & 0.4\% \\
Electron identification  & 100\% & 0\% & 1.2--2.2\% \\
Muon identification, syst. component  & 100\% & 0\% & 0.3\% \\
Muon identification, stat. component  & 0\% & 0\% & 0.5\% \\
Muon isolation, syst. component  & 100\% & 0\% & $<$0.1\% \\
Muon isolation, stat. component & 0\% & 0\% & 0.2\% \\
Phase-space extrapolation of lepton isolation  & 100\% & 100\% & 0.5--1\% \\
Jet energy scale, year-to-year correlated part & 100\% & 0\% & 0.8\% \\
Jet energy scale, year-to-year uncorrelated part & 0\% & 0\% & 1.4\% \\
Parton flavor impact on jet energy scale & 100\% & 100\% & 1.1\% \\
\PQb tagging & 0\% & 0\%  & 2--4\% \\[\cmsTabSkip]
Matrix element scale $^\dag$ & 100\% & 100\% & 0.3--6\% \\
PDF+\alpS $^\dag$ & 100\% & 100\% & 0.1--0.4\% \\
Initial- \& final-state radiation scale $^\dag$ & 100\% & 100\% & 1--5\% \\
Top quark \pt $^\dag$ & 100\% & 100\% & 0.5--2.5\%\\
Matrix element to parton shower matching $^\dag$ & 100\% & 100\% & 0.7\%\\
Underlying event tune $^\dag$ & 100\% & 100\% & 0.2\% \\
Color reconnection $^\dag$ & 100\%  & 100\% & 0.3\% \\
Top quark mass $^\dag$ & 100\%  & 100\% & 0.5--3\% \\[\cmsTabSkip]
Single top quark cross section $^\dag$ & 100\%  & 100\% & 30\% \\
\ttV cross section $^\dag$ & 100\%  & 100\% & 20\% \\
Diboson cross section $^\dag$ & 100\%  & 100\% & 30\%\\
\Vjets cross section $^\dag$ & 100\%  & 100\% & 30\% \\
\ttbar cross section $^{\ast\dag}$ & 100\%  & 100\% & 4\% \\[\cmsTabSkip]
Single top quark time modulation $^\ast$ & 100\%  & 100\% & 2\% \\[\cmsTabSkip]
MC statistical uncertainty & 0\% & 100\% & 0.1--1\% \\
\end{tabular}}
\end{table*}

\section{\label{DifferentialMeas}Measurement of the normalized \texorpdfstring{\ttbar}{ttbar} differential cross section with sidereal time}

The normalized differential cross section for \ttbar production is measured by employing a profile likelihood template fit of the predictions from simulation to the data including all sidereal time bins, based on the LHC test statistic~\cite{CMS:NOTE-2011-005,EXT:Asymptotic-2011, CMS:2024onh}.
Since the SM predictions for the \ttbar cross section are identical in each particle-level bin $i$, and because unfolding for detector effects is not needed (particle- and reconstructed-level sidereal time are identical), the normalized differential cross section $\sigma_i/\sigavg$ in bin $i$ (where \sigavg is the averaged cross section over all sidereal time bins) reduces to $\mu_i/\muavg$, where $\mu_i$ is the ratio of the observed \ttbar yield to the \ttbar yield predicted in the SM, and $\muavg = \sum_k \mu_k /24$.
The fractions $f_i = \mu_i/\muavg$ for 23 sidereal time bins are directly fitted as parameters of interest together with \muavg; the remaining fraction and its uncertainty is computed via error propagation using the covariance matrix of the fit.

\begin{figure*}[t!]
\centering
\includegraphics[width=\textwidth]{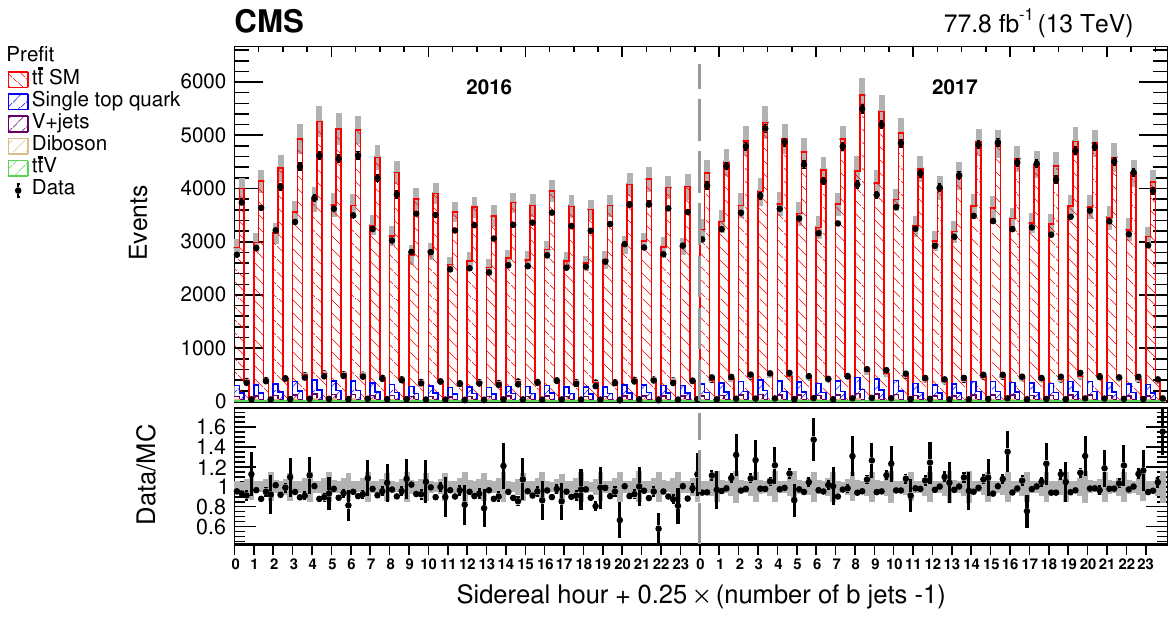}
\includegraphics[width=\textwidth]{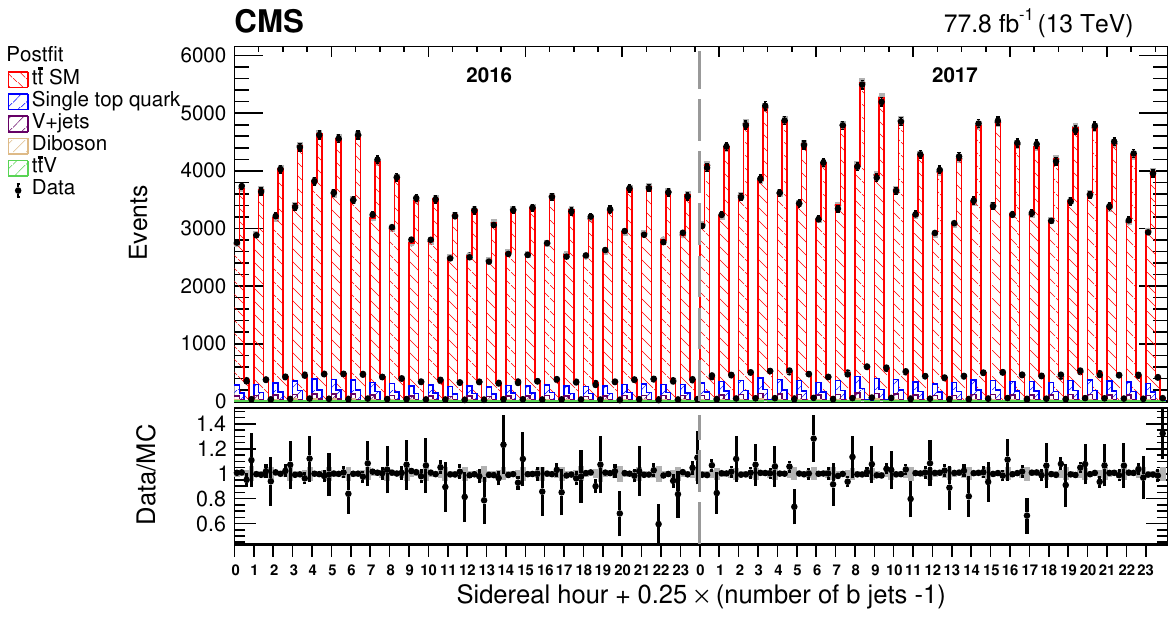}
\caption{\label{fig:n_bjets_prefit_differential} Prefit (upper) and postfit (lower) distributions of the number of \PQb jets in sidereal hour bins, in 2016 and 2017 data. The gray band reflects the statistical and systematic uncertainty predicted in each bin, including correlations across bins. The vertical bars associated with the data points represent their statistical uncertainty. The lower panels show the ratio of the observed data event yields to those expected from simulation.}
\end{figure*}

In performing the measurement of the \ttbar normalized differential cross section as a function of sidereal time, the distributions in the number of \PQb jets in each sidereal time bin are used as discriminant observables between \ttbar and background processes.
Uncertainties described in Section~\ref{Systematics} are considered either as nuisance parameters with log-normal distribution attributed to the normalization of the distribution in the number of \PQb jets, or as nuisance parameters distorting the shape of the distribution in the number of \PQb jets.
Uncertainties follow the correlation scheme highlighted in Table~\ref{CorrelationScheme}:\ when the dependency with sidereal time is known, the uncertainties are treated as correlated across sidereal time bin (including uniform theory and background normalization uncertainties); and when the dependency is not known, the uncertainties are treated as uncorrelated across sidereal time bin.
Prefit and postfit distributions are illustrated in Fig.~\ref{fig:n_bjets_prefit_differential}. The prefit distribution corresponds to the expected SM event yields and their uncertainties before the fit is performed. After the model is fitted to the data, the final values of the parameters and their uncertainties are obtained, corresponding to the postfit distribution.
The predicted distributions of the number of \PQb jets are not the same in each time bin, because corrections for integrated luminosity, pileup, and trigger efficiencies are applied as a function of sidereal time.

\begin{figure}[t!]
\centering
\ifthenelse{\boolean{cms@external}}{%
    \includegraphics[width=\linewidth]{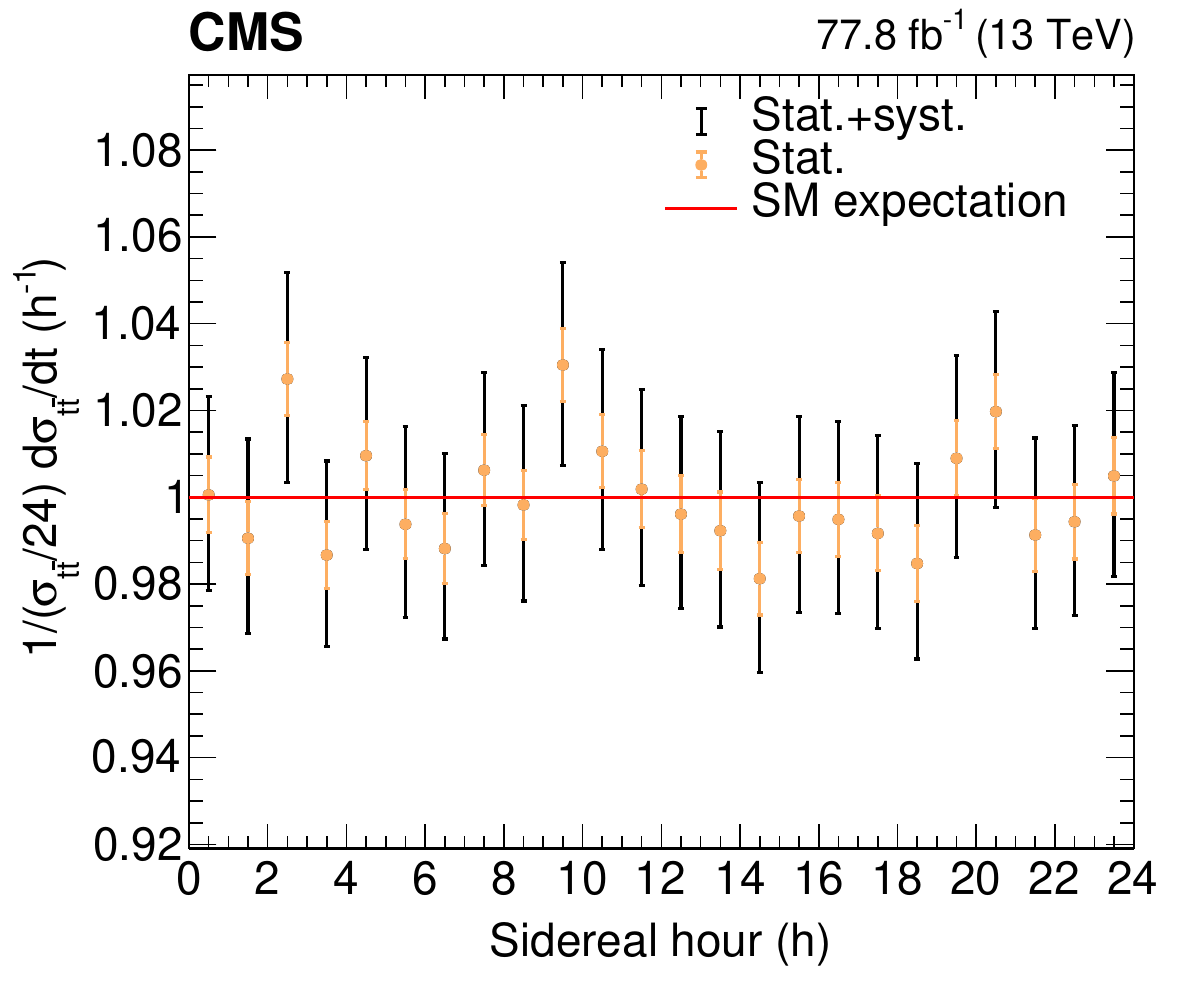}
}{%
    \includegraphics[width=0.6\textwidth]{Figure_003.pdf}
}
\caption{\label{fig:n_bjets_differential_yield} The normalized differential cross section for \ttbar as a function of sidereal time, using combined 2016--2017 data. The error bars show statistical, as well as statistical and systematic uncertainties, including correlations across bins.}
\end{figure}

The normalized differential cross section is shown in Fig.~\ref{fig:n_bjets_differential_yield}.
A goodness-of-fit test including constraints from the nuisance parameters~\cite{Baker:1983tu} results in a $p$-value~\cite{pvalue} of 0.92 for the adequacy of the MC model to describe data.

\begin{figure}[t!]
\centering
\ifthenelse{\boolean{cms@external}}{%
    \includegraphics[width=\linewidth]{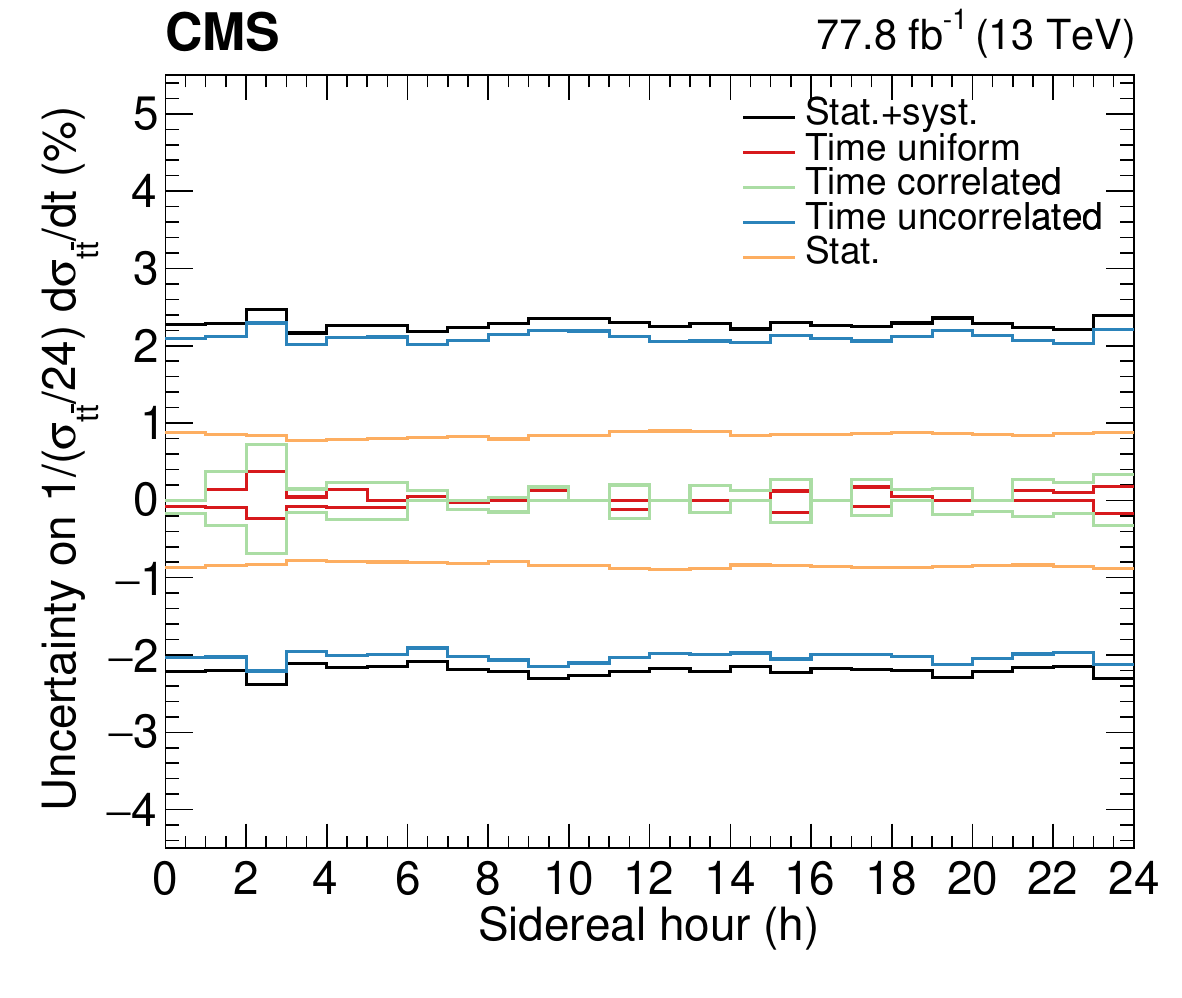}
}{%
    \includegraphics[width=0.6\textwidth]{Figure_004.pdf}
}
\caption{\label{fig:n_bjets_differential_yield_breakdown} Comparison of systematic and statistical uncertainties, where the former are grouped according to the treatment of time dependence:\ uniform (flat luminosity component, background normalization, theory), correlated (trigger, luminosity stability and linearity, pileup, and MC statistical uncertainty), or uncorrelated (other experimental uncertainties) across sidereal time bins.}
\end{figure}

Figure~\ref{fig:n_bjets_differential_yield_breakdown} shows a breakdown of the uncertainties associated with the measurement of \ttbar normalized differential cross section, according to the treatment of their dependences on sidereal time.
The measurement is dominated by systematic uncertainties.
The experimental uncertainties that are treated as uncorrelated in sidereal time have the largest impact on the normalized differential cross section, and they arise mainly from sources related to electrons, jet energy scale, and \PQb tagging.
The uncertainties featuring explicit dependence on sidereal time have an impact that is smaller than the statistical uncertainty, and are dominated by trigger uncertainties.
Uncertainties that are uniform in sidereal time (flat luminosity component, background normalization, theory uncertainties) contribute to a lesser extent.

\section{\label{SignalExtraction}Testing Lorentz invariance with the SME}

Corrections to the matrix element of \ttbar production and decay arising from Lorentz-violating terms in the SME are known at LO in pQCD~\cite{Berger:2015yha}, and are linear in the SME coefficients, stemming from the interference between SME and SM predictions.
Within the CMS reference frame, the variation of the \ttbar cross section with sidereal time $t$ is quantified with the function~\cite{Abazov:2012iu} $f(t) = \sigSME(t)/\sigSM-1$, where \sigSME (\sigSM) are the values predicted in the SME (SM):
\ifthenelse{\boolean{cms@external}}{\begin{multline}
    \label{eqFt2}
    f(t) = (\cLmunu+\cRmunu) R^{\mu}_{\alpha}(t) R^{\nu}_{\beta}(t) \Bigg(\frac{\delta P}{P} \Bigg) ^{\alpha\beta} \\
    + \cLmunu R^{\mu}_{\alpha}(t) R^{\nu}_{\beta}(t) \Bigg( \frac{\delta F}{F} + \frac{\delta \bar{F}}{\bar{F}} \Bigg)^{\alpha\beta},
\end{multline}}{\begin{equation}
    \label{eqFt2}
    f(t) = (\cLmunu+\cRmunu) R^{\mu}_{\alpha}(t) R^{\nu}_{\beta}(t) \Bigg(\frac{\delta P}{P} \Bigg) ^{\alpha\beta}
    + \cLmunu R^{\mu}_{\alpha}(t) R^{\nu}_{\beta}(t) \Bigg( \frac{\delta F}{F} + \frac{\delta \bar{F}}{\bar{F}} \Bigg)^{\alpha\beta},
\end{equation}}
with $\cLmunu = \cmunu+\dmunu$ and $\cRmunu =  \cmunu-\dmunu$; where $\alpha$ and $\beta$ represent summation over space-time indices in the CMS detector frame; $P$, $F$, and $\bar{F}$ are the squared LO matrix elements of the SM \ttbar production (either quark-antiquark annihilation or gluon fusion), and top quark and antiquark decays; and $\delta P$, $\delta F$, $\delta \bar{F}$ are their corrections predicted in the SME.
Rotation matrices $R(t)$ represent transformations of observer coordinates from the SCF to the CMS reference frame, depending on the Earth's rotation around its axis with angular velocity \Omgsid.
The coordinates of the CMS interaction point are specified~\cite{CERN-WGS84} by the azimuth on the LHC ring ($\phi = 101.28^{\circ}$), the latitude ($\lambda = 46.31^{\circ}$~N), and the longitude ($\ell = 6.08^{\circ}$~E).
Furthermore, the LHC plane, and thus the CMS cavern, has an angle of $\gamma = 0.705^{\circ}$ relative to the surface.

A \ttbar production sample, generated with \MGvATNLO at LO in pQCD with parton shower, underlying event, and hadronization simulated with \PYTHIA8, followed by full detector simulation, is used to evaluate the signal $f(t)$ functions.
The $f(t)$ functions are computed for each number of reconstructed \PQb jets separately for each year and sidereal time bin.
Events containing exactly 1, 2, 3, or $\geq$4 \PQb jets at the reconstructed level are used to evaluate the matrix elements $P$, $F$, $\bar{F}$ and their SME corrections, using exact LO kinematics before showering.

\begin{figure*}[htp!]
\centering
\includegraphics[width=0.95\textwidth]{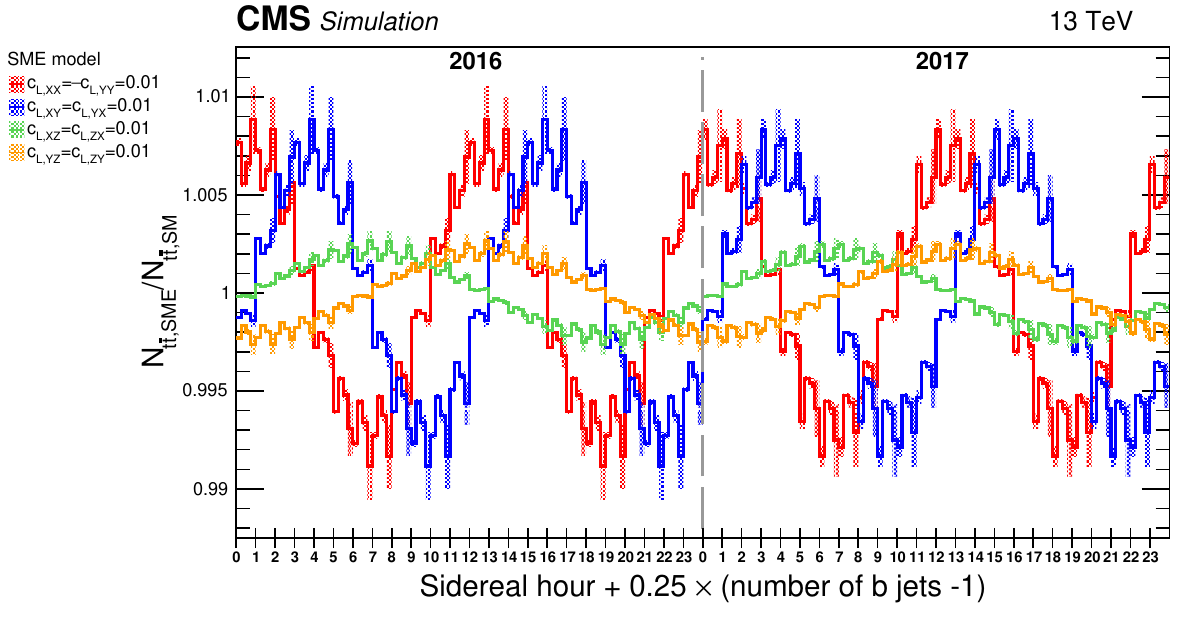}
\caption{\label{fig:f_function_cmunu_n_bjets_reco_unrolled} Number of \ttbar events reconstructed in the SME hypothesis divided by the number of events in the SM hypothesis, as a function of the number of \PQb jets and sidereal time, for the four directions of the $c_L$ coefficients. The uncertainty band represents the MC statistical uncertainty in the sample used to compute the SME hypothesis. The sinusoidal variation is arising from the $f(t)$ dependence on sidereal time, while smaller structures reflect the number of b jets in each sidereal time bin.}
\end{figure*}

The SME coefficients are classified into four families extracted separately, corresponding to \cLmunu (assuming $\cRmunu=0$), \cRmunu (assuming $\cLmunu=0$), \cmunu (assuming $\dmunu = 0$), and \dmunu (assuming $\cmunu = 0$).
Within the chosen basis of coefficients adopted in the SME, these matrices are defined as symmetric and traceless.
Coefficients with indices $\mu\nu=TT$ induce a shift in the \ttbar cross section and are not considered.
Since the $Z$-axis is defined as the Earth's rotation axis, and because modulation of the \ttbar cross section with sidereal time is induced by rotation around this axis, there is by definition no sensitivity in the coefficients with indices $\mu\nu=ZZ$.
For similar reasons, coefficients with indices $\mu\nu=ZT$ are not considered.
Coefficients with $\mu\nu=XT$ or $YT$ are found to induce very small SME corrections and are also not considered.
Remaining coefficients correspond to the combinations $c_{XX}=-c_{YY}$, $c_{XZ}=c_{ZX}$, $c_{YZ}=c_{ZY}$, and $c_{XY}=c_{YX}$ (and similar expressions for the three other scenarios).
A total of 16 sets of coefficients is determined.
An illustration of their impact on the \ttbar cross section is shown in Fig.~\ref{fig:f_function_cmunu_n_bjets_reco_unrolled}.

\begin{figure*}[bp!]
\centering
\includegraphics[width=0.75\textwidth]{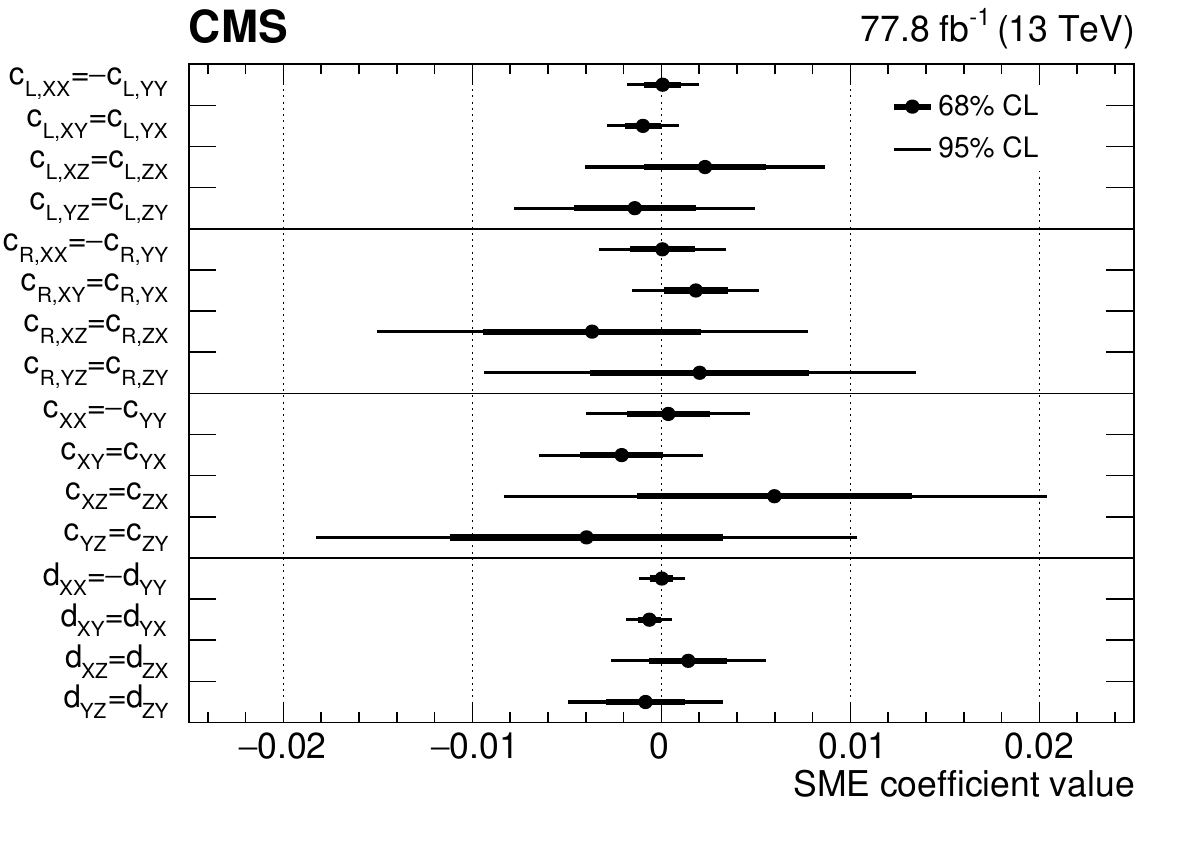}
	\caption{\label{fig:SMEwilsonPlot_Comb_data}Fitted SME coefficients and their 68 and 95\% CL, measured in fits of single coefficients while the coefficients corresponding to the three other directions are left floating, within the $c_L$, $c_R$, $c$, and $d$ families. The error bar includes statistical and systematic uncertainties. Fitting a single coefficient, with the others fixed to the SM value, leads to negligible changes in the results.}
\end{figure*}

The Lagrangian defined in Eq.~\eqref{eqFt1} impacts all processes involving top quarks, including single top quark and \ttV processes.
The latter processes have a small yield and their variation with sidereal time can be neglected, contrary to the former processes.
Expressions for the matrix elements of single top quark production in the SME are not available, to our knowledge.
Among the single top quark processes, \tW production has the largest contribution in the selected dileptonic event sample.
Therefore, uncertainties corresponding to single top quark time modulation in the SME are attributed to single top quark processes, and computed from matrix elements predicted in the SME for top quark and antiquark decay only, in an MC sample of \tW processes generated at LO.
These uncertainties correspond to a value of the SME coefficients of $\pm 3\times 10^{-3}$ in $XX$ and $XY$ directions, and $\pm 10\times 10^{-3}$ in $XZ$ and $YZ$ directions, at the same order of magnitude as the expected sensitivity in the worst case among $c_L$, $c_R$, $c$, and $d$ sets of coefficients.
In the case of \cRmunu coefficients, which do not affect the top quark or antiquark in the decay (as highlighted in Eq.~\eqref{eqFt2}), the uncertainty model for \cLmunu is used.
In measuring individually the SME coefficients, the SME uncertainty for single top quark processes in a given direction has been verified to impact only the coefficients related to the same direction, with an uncertainty corresponding to approximately 15\% of the total uncertainty on the coefficient.

\begin{table*}[htp!]
\centering
\topcaption{\label{SMEresultsAsimDataSingleMultiple}Expected and observed 68\% confidence level interval measured for the SME fits of single coefficients while the others are fixed to their SM value, and while coefficients for the three other directions are floating. The expected intervals represent the results obtained using simulation only, assuming null values of the coefficients in the data. Reported intervals are to be multiplied by $10^{-3}$.}
\renewcommand{\arraystretch}{1.1}
\begin{tabular}{>{\hspace{1.5ex}}r@{${}={}$}l>{[}r@{; }l<{]}>{[}r@{; }l<{]}>{[}r@{; }l<{]}>{[}r@{; }l<{]}}
\multicolumn{2}{c}{SME coefficient} & \multicolumn{4}{c}{Others fixed to SM} & \multicolumn{4}{c}{Others floating} \\
	\multicolumn{2}{c}{($10^{-3}$ unit)} & \multicolumn{2}{c}{Expected} & \multicolumn{2}{c}{Observed} & \multicolumn{2}{c}{Expected} & \multicolumn{2}{c}{Observed} \\
\hline
$c_{L,XX}$ & $-c_{L,YY}$ & $-0.96$ & $0.96$ & $-0.9$ & $1.03$ & $-0.96$ & $0.96$ & $-0.9$ & $1.03$ \\
$c_{L,XY}$ & $c_{L,YX}$ & $-0.97$ & $0.97$ & $-1.92$ & $0.0$ & $-0.97$ & $0.97$ & $-1.94$ & $-0.02$ \\
$c_{L,XZ}$ & $c_{L,ZX}$ & $-3.23$ & $3.23$ & $-0.97$ & $5.49$ & $-3.23$ & $3.23$ & $-0.92$ & $5.54$ \\
$c_{L,YZ}$ & $c_{L,ZY}$ & $-3.24$ & $3.24$ & $-4.61$ & $1.85$ & $-3.24$ & $3.24$ & $-4.64$ & $1.82$ \\[\cmsTabSkip]
$c_{R,XX}$ & $-c_{R,YY}$ & $-1.7$ & $1.7$ & $-1.65$ & $1.77$ & $-1.7$ & $1.7$ & $-1.66$ & $1.76$ \\
$c_{R,XY}$ & $c_{R,YX}$ & $-1.71$ & $1.71$ & $0.09$ & $3.5$ & $-1.71$ & $1.71$ & $0.12$ & $3.52$ \\
$c_{R,XZ}$ & $c_{R,ZX}$ & $-5.78$ & $5.78$ & $-9.36$ & $2.2$ & $-5.78$ & $5.78$ & $-9.45$ & $2.11$ \\
$c_{R,YZ}$ & $c_{R,ZY}$ & $-5.8$ & $5.8$ & $-3.82$ & $7.76$ & $-5.8$ & $5.8$ & $-3.77$ & $7.82$ \\[\cmsTabSkip]
$c_{XX}$ & $-c_{YY}$ & $-2.17$ & $2.17$ & $-1.76$ & $2.62$ & $-2.17$ & $2.17$ & $-1.83$ & $2.55$ \\
$c_{XY}$ & $c_{YX}$ & $-2.18$ & $2.18$ & $-4.23$ & $0.17$ & $-2.18$ & $2.18$ & $-4.31$ & $0.09$ \\
$c_{XZ}$ & $c_{ZX}$ & $-7.21$ & $7.21$ & $-1.49$ & $13.07$ & $-7.21$ & $7.21$ & $-1.29$ & $13.27$ \\
$c_{YZ}$ & $c_{ZY}$ & $-7.24$ & $7.24$ & $-11.05$ & $3.38$ & $-7.24$ & $7.24$ & $-11.21$ & $3.28$ \\[\cmsTabSkip]
$d_{XX}$ & $-d_{YY}$ & $-0.61$ & $0.61$ & $-0.6$ & $0.63$ & $-0.61$ & $0.61$ & $-0.59$ & $0.64$ \\
$d_{XY}$ & $d_{YX}$ & $-0.62$ & $0.62$ & $-1.24$ & $-0.01$ & $-0.62$ & $0.62$ & $-1.25$ & $-0.02$ \\
$d_{XZ}$ & $d_{ZX}$ & $-2.07$ & $2.07$ & $-0.68$ & $3.46$ & $-2.08$ & $2.07$ & $-0.65$ & $3.49$ \\
$d_{YZ}$ & $d_{ZY}$ & $-2.08$ & $2.08$ & $-2.9$ & $1.25$ & $-2.08$ & $2.08$ & $-2.92$ & $1.23$ \\
\end{tabular}
\end{table*}

The discriminant observable used to extract the SME coefficients is the same as in the differential measurement (Section~\ref{DifferentialMeas}), with event corrections and systematic uncertainties described in Section~\ref{Systematics}.
The likelihood fit is performed in two scenarios:\ 1) independently for each set of coefficients as a single parameter of interest, with all other parameters being set to zero; 2) independently for each set of coefficients as a single parameter of interest, with the parameters corresponding to the three other directions (among $XX$, $XY$, $XZ$, $YZ$ directions) floating in the fit, for the four families of coefficients $c_L$, $c_R$, $c$, and $d$ separately.
The measured values of the SME sets of coefficients in the latter scenario is shown in Fig.~\ref{fig:SMEwilsonPlot_Comb_data}, while Table~\ref{SMEresultsAsimDataSingleMultiple} compares the values for the two scenarios.
Scenarios 1 and 2 yield expected uncertainties identical up to the second digit, because the correlation between the four sets of coefficients in a family is found to be 0--4\%.
A $p$-value of 0.98 is obtained for the measurements of the coefficients in a given direction while the coefficients corresponding to the three other directions are floating in the fit.
All of the measured values for the SME coefficients are compatible with zero, in agreement with the SM within one standard deviation or less, indicating no sign of deviation from Lorentz invariance.
The precision achieved on the values of the coefficients of the $c_L$ and $c_R$ families on the one hand, and on the values of the coefficients of the $d$ family on the other hand, is improved respectively by a factor 20--50 and up to 100 over the previous constraints~\cite{Abazov:2012iu}. The coefficients of the $c$ family are measured for the first time.

\begin{figure*}[htp!]
\centering
\includegraphics[width=0.6\textwidth]{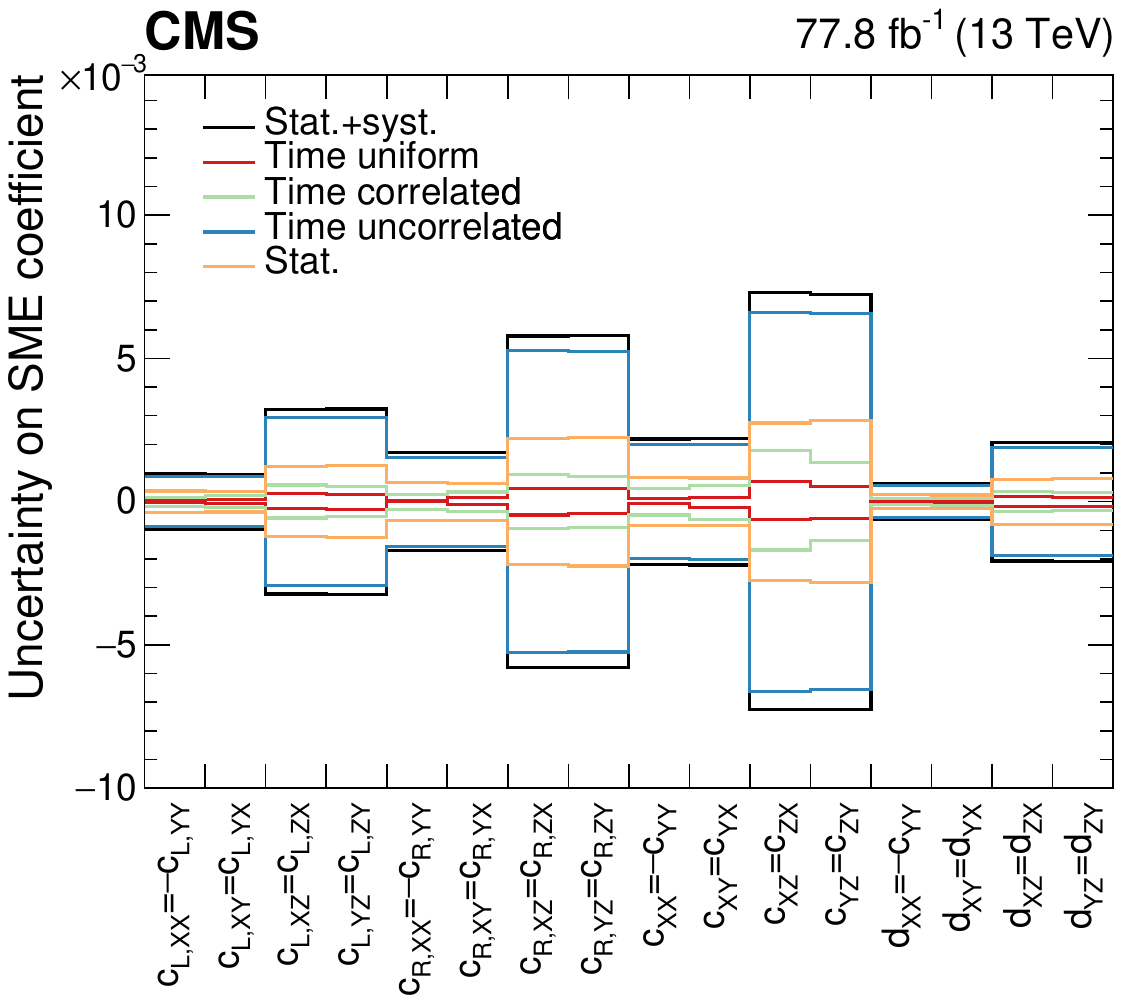}
\caption{\label{fig:n_bjets_smefit_breakdown_time_data} Uncertainty breakdown for SME fits of single coefficients while the coefficients corresponding to the three other directions are left floating, by splitting according to the treatment of time dependence:\ flat across sidereal time (flat luminosity component, background normalization, theory), correlated in sidereal time bins (trigger, luminosity stability and linearity, pileup, MC statistical uncertainty, single top quark decay in the SME), systematics uncorrelated in sidereal time bins (other experimental uncertainties), and statistical uncertainty.}
\end{figure*}

The impact of the uncertainties in the SME coefficients is shown in Fig.~\ref{fig:n_bjets_smefit_breakdown_time_data} in the case of the fit of individual sets of coefficients, while the coefficients corresponding to the three other directions are left floating.
Systematic uncertainties uncorrelated between sidereal time bins are dominating the precision, at a level of approximately three times higher than the statistical uncertainties.
The systematic uncertainties that are nonuniform and correlated across sidereal time impact the precision at about half the impact of the statistical uncertainties.
The uncertainties that are uniform in sidereal time have a smaller impact.

\section{\label{Conclusions}Summary}

A search for violation of Lorentz invariance has been performed using top quark pairs (\ttbar), requiring the presence of one muon and one electron in the events.
Data collected in 2016--2017 with the CMS detector corresponding to an integrated luminosity of 77.8\fbinv are used.
A measurement of the \ttbar normalized differential cross section as a function of sidereal time is performed.
The Lorentz invariance assumption is tested by measuring 16 sets of coefficients within the standard model extension, an effective field theory predicting a modulation of the \ttbar cross section with sidereal time.
Measurements of the Lorentz-violating couplings are found to be compatible with the standard model hypothesis.
The precision of the results ranges from less than $1\times 10^{-3}$ to $8\times 10^{-3}$ for the measured coefficients.
This constitutes the most precise test of the isotropy in special relativity using top quarks at a hadron collider.

\begin{acknowledgments}
We congratulate our colleagues in the CERN accelerator departments for the excellent performance of the LHC and thank the technical and administrative staffs at CERN and at other CMS institutes for their contributions to the success of the CMS effort. In addition, we gratefully acknowledge the computing centers and personnel of the Worldwide LHC Computing Grid and other centers for delivering so effectively the computing infrastructure essential to our analyses. Finally, we acknowledge the enduring support for the construction and operation of the LHC, the CMS detector, and the supporting computing infrastructure provided by the following funding agencies: SC (Armenia), BMBWF and FWF (Austria); FNRS and FWO (Belgium); CNPq, CAPES, FAPERJ, FAPERGS, and FAPESP (Brazil); MES and BNSF (Bulgaria); CERN; CAS, MoST, and NSFC (China); MINCIENCIAS (Colombia); MSES and CSF (Croatia); RIF (Cyprus); SENESCYT (Ecuador); ERC PRG, RVTT3 and MoER TK202 (Estonia); Academy of Finland, MEC, and HIP (Finland); CEA and CNRS/IN2P3 (France); SRNSF (Georgia); BMBF, DFG, and HGF (Germany); GSRI (Greece); NKFIH (Hungary); DAE and DST (India); IPM (Iran); SFI (Ireland); INFN (Italy); MSIP and NRF (Republic of Korea); MES (Latvia); LMTLT (Lithuania); MOE and UM (Malaysia); BUAP, CINVESTAV, CONACYT, LNS, SEP, and UASLP-FAI (Mexico); MOS (Montenegro); MBIE (New Zealand); PAEC (Pakistan); MES and NSC (Poland); FCT (Portugal); MESTD (Serbia); MCIN/AEI and PCTI (Spain); MOSTR (Sri Lanka); Swiss Funding Agencies (Switzerland); MST (Taipei); MHESI and NSTDA (Thailand); TUBITAK and TENMAK (Turkey); NASU (Ukraine); STFC (United Kingdom); DOE and NSF (USA).

\hyphenation{Rachada-pisek} Individuals have received support from the Marie-Curie program and the European Research Council and Horizon 2020 Grant, contract Nos.\ 675440, 724704, 752730, 758316, 765710, 824093, 101115353, 101002207, and COST Action CA16108 (European Union); the Leventis Foundation; the Alfred P.\ Sloan Foundation; the Alexander von Humboldt Foundation; the Science Committee, project no. 22rl-037 (Armenia); the Belgian Federal Science Policy Office; the Fonds pour la Formation \`a la Recherche dans l'Industrie et dans l'Agriculture (FRIA-Belgium); the Agentschap voor Innovatie door Wetenschap en Technologie (IWT-Belgium); the F.R.S.-FNRS and FWO (Belgium) under the ``Excellence of Science -- EOS" -- be.h project n.\ 30820817; the Beijing Municipal Science \& Technology Commission, No. Z191100007219010 and Fundamental Research Funds for the Central Universities (China); the Ministry of Education, Youth and Sports (MEYS) of the Czech Republic; the Shota Rustaveli National Science Foundation, grant FR-22-985 (Georgia); the Deutsche Forschungsgemeinschaft (DFG), under Germany's Excellence Strategy -- EXC 2121 ``Quantum Universe" -- 390833306, and under project number 400140256 - GRK2497; the Hellenic Foundation for Research and Innovation (HFRI), Project Number 2288 (Greece); the Hungarian Academy of Sciences, the New National Excellence Program - \'UNKP, the NKFIH research grants K 131991, K 133046, K 138136, K 143460, K 143477, K 146913, K 146914, K 147048, 2020-2.2.1-ED-2021-00181, and TKP2021-NKTA-64 (Hungary); the Council of Science and Industrial Research, India; ICSC -- National Research Center for High Performance Computing, Big Data and Quantum Computing and FAIR -- Future Artificial Intelligence Research, funded by the NextGenerationEU program (Italy); the Latvian Council of Science; the Ministry of Education and Science, project no. 2022/WK/14, and the National Science Center, contracts Opus 2021/41/B/ST2/01369 and 2021/43/B/ST2/01552 (Poland); the Funda\c{c}\~ao para a Ci\^encia e a Tecnologia, grant CEECIND/01334/2018 (Portugal); the National Priorities Research Program by Qatar National Research Fund; MCIN/AEI/10.13039/501100011033, ERDF ``a way of making Europe", and the Programa Estatal de Fomento de la Investigaci{\'o}n Cient{\'i}fica y T{\'e}cnica de Excelencia Mar\'{\i}a de Maeztu, grant MDM-2017-0765 and Programa Severo Ochoa del Principado de Asturias (Spain); the Chulalongkorn Academic into Its 2nd Century Project Advancement Project, and the National Science, Research and Innovation Fund via the Program Management Unit for Human Resources \& Institutional Development, Research and Innovation, grant B37G660013 (Thailand); the Kavli Foundation; the Nvidia Corporation; the SuperMicro Corporation; the Welch Foundation, contract C-1845; and the Weston Havens Foundation (USA).
\end{acknowledgments}

\bibliography{auto_generated}
\cleardoublepage \appendix\section{The CMS Collaboration \label{app:collab}}\begin{sloppypar}\hyphenpenalty=5000\widowpenalty=500\clubpenalty=5000
\cmsinstitute{Yerevan Physics Institute, Yerevan, Armenia}
{\tolerance=6000
A.~Hayrapetyan, A.~Tumasyan\cmsAuthorMark{1}\cmsorcid{0009-0000-0684-6742}
\par}
\cmsinstitute{Institut f\"{u}r Hochenergiephysik, Vienna, Austria}
{\tolerance=6000
W.~Adam\cmsorcid{0000-0001-9099-4341}, J.W.~Andrejkovic, T.~Bergauer\cmsorcid{0000-0002-5786-0293}, S.~Chatterjee\cmsorcid{0000-0003-2660-0349}, K.~Damanakis\cmsorcid{0000-0001-5389-2872}, M.~Dragicevic\cmsorcid{0000-0003-1967-6783}, P.S.~Hussain\cmsorcid{0000-0002-4825-5278}, M.~Jeitler\cmsAuthorMark{2}\cmsorcid{0000-0002-5141-9560}, N.~Krammer\cmsorcid{0000-0002-0548-0985}, A.~Li\cmsorcid{0000-0002-4547-116X}, D.~Liko\cmsorcid{0000-0002-3380-473X}, I.~Mikulec\cmsorcid{0000-0003-0385-2746}, J.~Schieck\cmsAuthorMark{2}\cmsorcid{0000-0002-1058-8093}, R.~Sch\"{o}fbeck\cmsorcid{0000-0002-2332-8784}, D.~Schwarz\cmsorcid{0000-0002-3821-7331}, M.~Sonawane\cmsorcid{0000-0003-0510-7010}, S.~Templ\cmsorcid{0000-0003-3137-5692}, W.~Waltenberger\cmsorcid{0000-0002-6215-7228}, C.-E.~Wulz\cmsAuthorMark{2}\cmsorcid{0000-0001-9226-5812}
% ...
\par}
\vskip\cmsinstskip
\dag:~Deceased\\
$^{1}$Also at Yerevan State University, Yerevan, Armenia\\
$^{2}$Also at TU Wien, Vienna, Austria\\
% ...
\end{sloppypar}
%%% END EDITABLE REGION %%%
% skeleton_end
\end{document}